\documentclass{jpsj3}
\usepackage{txfonts}
\usepackage{color}

\title{Ferroelectric Transition of a Chiral Molecular Crystal BINOL$\cdot$2DMSO}

\author{Toshihiro Nomura$^1$\thanks{t.nomura@issp.u-tokyo.ac.jp},
 Takeshi Yajima$^1$, Zhuo Yang$^1$, Ryosuke Kurihara$^1$, Yuto Ishii$^1$, Masashi Tokunaga$^1$, Yasuhiro H. Matsuda$^1$, Yoshimitsu Kohama$^1$, Kenta Kimura$^2$, and Tsuyoshi Kimura$^2$}
\inst{$^1$Institute for Solid State Physics, University of Tokyo, Kashiwa, Chiba 277-8581, Japan\\
$^2$Department of Advanced Materials Science, University of Tokyo, Kashiwa, Chiba 277-8561, Japan}

\abst{We report dielectric, thermodynamical, acoustic, and optical properties of a chiral molecular crystal, 1,1’-bi-2-naphthol 2-dimethylsulfoxide (BINOL$\cdot$2DMSO). 
We find two successive phase transitions at $T_\mathrm{c1}=190$ K and $T_\mathrm{c2}=125$ K. 
The first transition at $T_\mathrm{c1}$ is characterized by an order-disorder transition of the guest molecules DMSO along with ferroelectricity. 
At the second transition of $T_\mathrm{c2}$, the crystal structure deforms from tetragonal to monoclinic, leading to domain formation. 
Low-temperature x-ray diffraction suggests that the space group changes from $P4_12_12$ ($P4_32_12$) to $P4_1$ ($P4_3$) at $T_\mathrm{c1}$, and down to $P112_1$ at $T_\mathrm{c2}$.
}

\begin{document}
\maketitle

\section{Introduction}
Ferroelectricity was first discovered in Rochelle salt (KNaC$_4$H$_4$O$_6\cdot$4H$_2$O),
a chiral metal-organic compound \cite{PhysRev.17.475,doi:10.1098/rspa.1941.0010,PhysRevB.53.5217,Mo2015}. 
Indeed, chirality and ferroelectricity are closely related in symmetry; mirror symmetry breaks in the former, while inversion symmetry breaks in the latter.
Chiral molecules are easier to crystallize in ferroelectric structures and can serve as an attractive system for the search of functional materials \cite{Li5878}.

\begin{figure}[p]
\centering
\includegraphics[width=7.5cm]{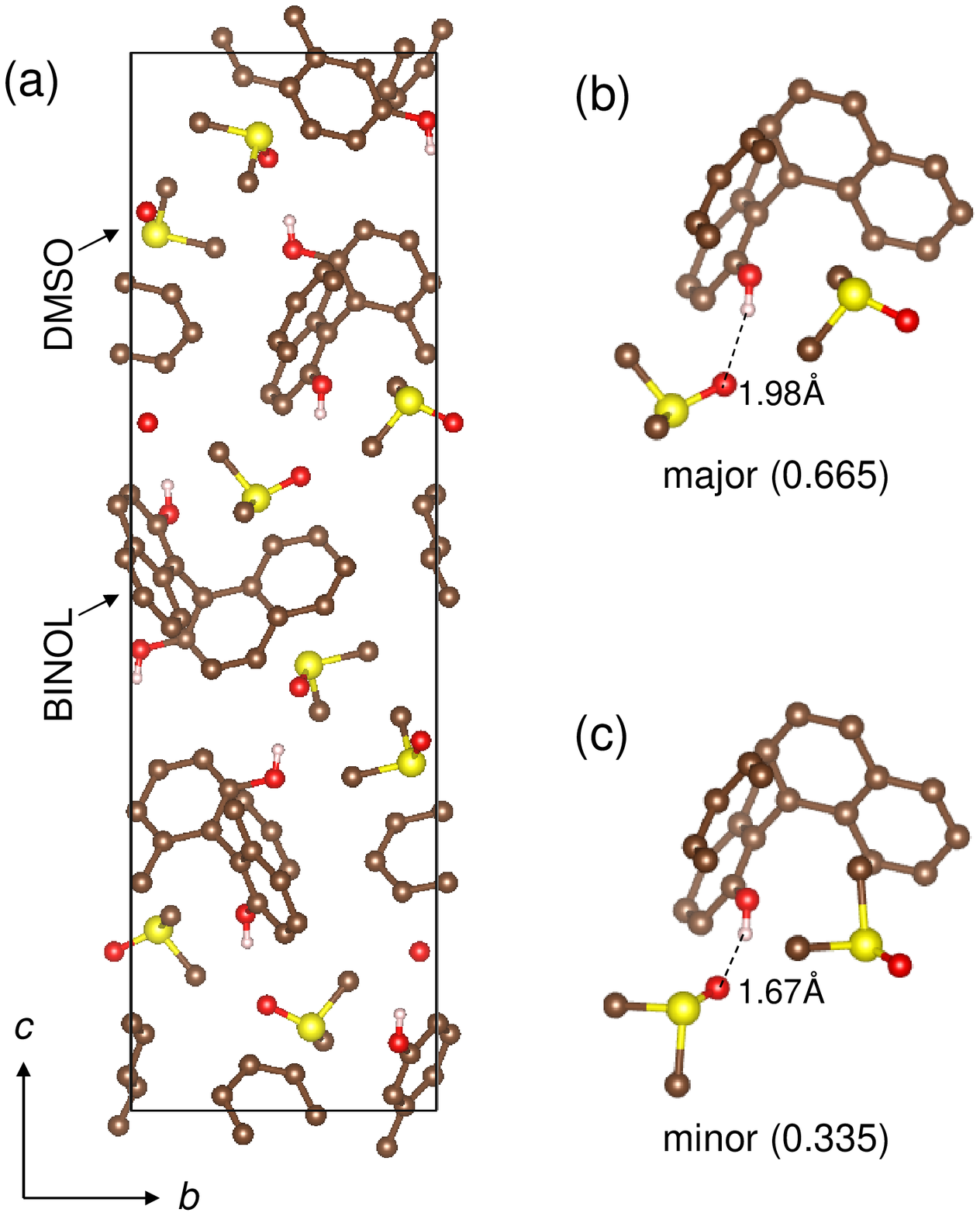}
\caption{\label{fig:crystal}
(Color online) (a) Crystal structure of BINOL$\cdot$2DMSO at room temperature projected along the $a$ axis \cite{doi:10.1021/cg1004648}. 
Brown, red, white, and yellow balls represent C, O, H, and S atoms, respectively.
C-H bonding is omitted for clarity.
(b, c) Enlarged view of the hydrogen bonding between BINOL and DMSO molecules.
(b) Major and (c) minor configurations of DMSO.
In Fig. \ref{fig:crystal}(a), the DMSO molecules are shown as the major configuration, although these two configurations fluctuate in reality.
The figures are drawn by using VESTA software \cite{Momma:ko5060}.
}
\end{figure}

In this study, we focus on the physical properties of a chiral molecule 1,1’-bi-2-naphthol (BINOL), where two enantiomers are distinguished by the axial chirality \cite{https://doi.org/10.1002/cphc.201800950, molecules21111541}. 
By using enantiomerically-pure BINOL, the resultant crystal structure also becomes right- or left-handed depending on the handedness of BINOL. 
Some organic crystals composed of BINOL molecules are reported by Lee and Peng \cite{doi:10.1021/cg1004648}. 
Among them, we pick up an enantiomorphic system (R)-(+)-BINOL$\cdot$2DMSO [(S)-(-)-BINOL$\cdot$2DMSO], where the crystal contains two dimethylsulfoxide (DMSO) molecules as guests [Fig. \ref{fig:crystal}(a)]. 
Here, the alignments of the guest DMSO molecules are partially disordered in the chiral framework of BINOL [Fig. \ref{fig:crystal}(b, c)]. 
Since a DMSO has a large dipole moment of 3.96D \cite{NSRDS}, the dipole-dipole interactions (and/or hydrogen bonding with BINOL) can result in an order-disorder transition of DMSO at low temperatures. 
Besides, the ordering of the polar molecules DMSO indicates ferroelectricity or antiferroelectricity in the ordered phase. 
Recently, organic ferroelectrics catch more attentions because of their environment-friendly features \cite{doi:10.1126/science.1229675,doi:10.1126/science.aas9330,Horiuchi2008}. 
In this paper, we report a ferroelectric transition of BINOL$\cdot$2DMSO studied by the dielectric, thermodynamical, acoustic, and optical techniques. 
Besides, the crystal structure at low temperature and the related ordering of DMSO molecules are discussed based on the low-temperature powder x-ray diffraction (XRD).

\section{Experiment}
We purchased enantiomerically-pure BINOL powder from Fuji Molecular Planning Co., Ltd. 
Single crystals of (R)-(+)-BINOL$\cdot$2DMSO [(S)-(-)-BINOL$\cdot$2DMSO] were grown by slow evaporation of the saturated DMSO solutions at room temperature. 
Transparent single crystals with the typical size of $1\times1\times1$ mm$^3$ were obtained in 2--3 days. 
We confirmed that the results of the dielectric measurements were the same for the right- and left-handed crystals. 
In the following, we do not distinguish the results for (R)-(+)-BINOL$\cdot$2DMSO or (S)-(-)-BINOL$\cdot$2DMSO.
In this paper, we do not discuss the properties of the racemic compound, which also shows dielectric anomalies at low temperatures.

We measured the dielectric properties on single crystals with applied electric fields along $E||c$ and $E||a$. 
We attached two gold wires to the as-grown surfaces by using Ag paint as electrodes with the typical area of $\sim0.5$~mm$^2$. 
The dielectric constant $\varepsilon^{\prime}$ and dissipation factor $D$ were measured by the LCR meter Keysight E4980A. 
The pyrocurrent ($I$) measurements were performed by using the electrometer Keithley 6517A/B. 
We measured a displacement current $I(t)$ induced by sweeping an applied voltage or temperature.
Polarization $P$ was obtained by numerically integrating $I(t)$.

We performed specific-heat measurements by using the relaxation technique of the Physical Property Measurement System (Quantum Design Ltd.). 
We used 3 mg of single crystals and attached them to the sample stage by using Apiezon N grease.

We performed sound-velocity measurements by using an ultrasound pulse-echo technique. 
We measured the longitudinal ($c_{33}$, $k||u||[001]$) and transverse ($c_{44}$,  $k||[001]$, $u||[100]$) mode by using Y-36$^{\circ}$- and X-41$^{\circ}$-cut LiNbO$_3$ transducers, respectively. 
We measured at the fundamental frequencies of the transducers, 38 MHz ($c_{33}$ mode) and 18 MHz ($c_{44}$ mode). 
We attached two transducers to the $c$ surfaces of the single crystal by using Thiokol.
Temperature dependence of the echo pattern was recorded by a digital storage oscilloscope (LeCroy HDO4104A). 
The phase of the 0th echo was numerically analyzed and the relative change of the sound velocity $\Delta v/v_0$ was obtained. 
The estimated error for the sound velocity was around 5\%, while the error for $\Delta v/v_0$ was around 1\%.

We performed optical absorption spectroscopy with incident light along the $c$ axis by using JASCO V-570. 
The sample thickness was 2.6 and 1.1 mm for the measurement run \#1 and \#2, respectively.
The sample was cooled in a He-gas-flow optical cryostat. 
In addition to the spectroscopy, we also observed by eyes how the outlook of the crystal changes at low temperatures.

We performed low-temperature powder XRD measurements by using an x-ray diffractometer (SmartLab Rigaku) with Cu $\rm{K\alpha}_{1}$ radiation monochromated by a Ge(111)-Johansson-type monochromator.
We ground single crystals of BINOL$\cdot$2DMSO to paste in a mortar.
Here, we added a small amount of DMSO to avoid deterioration because the BINOL$\cdot$2DMSO crystals can lose the guest DMSO molecules in a vacuum. 
We analyzed the XRD powder patterns by using PDXL2 software (Rigaku).

\section{Results}
\subsection{Dielectric properties}
\begin{figure}[p]
\centering
\includegraphics[width=9cm]{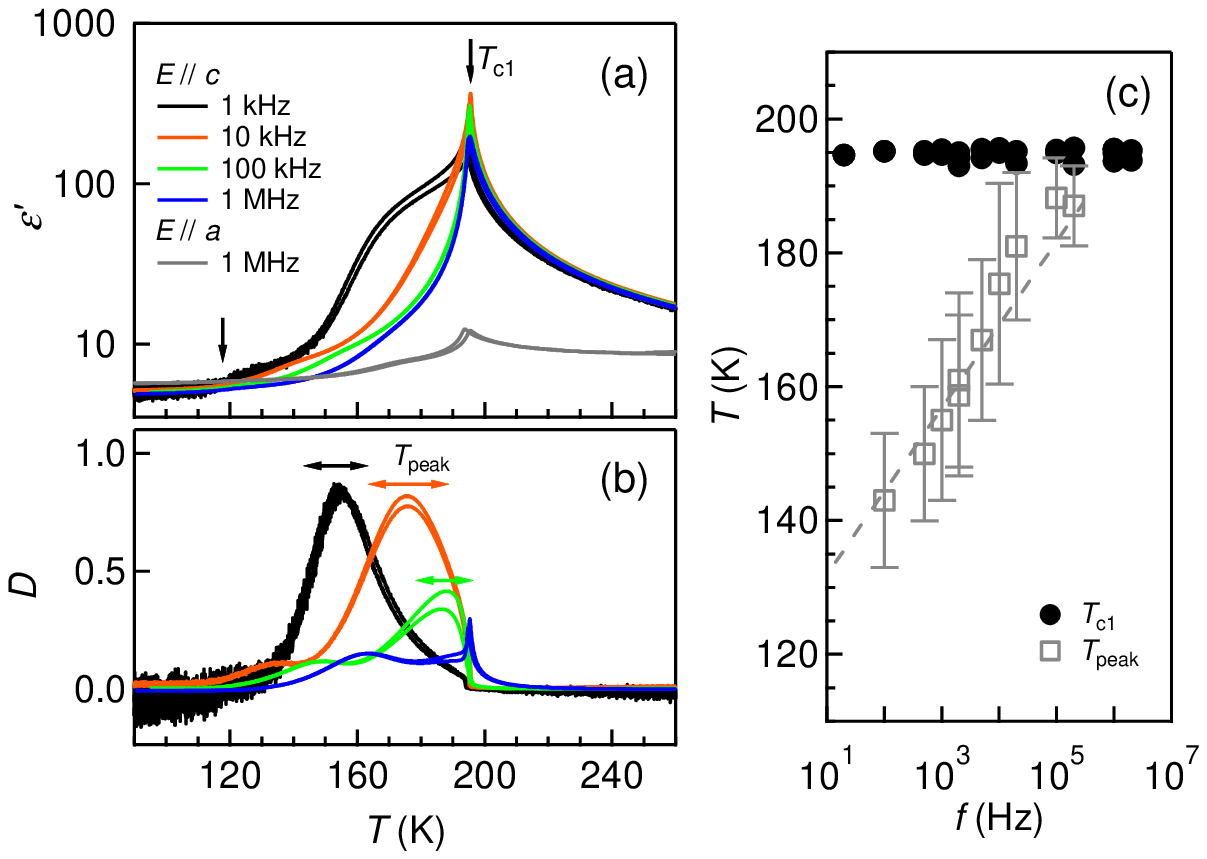}
\caption{\label{fig:eps}
(Color online) Temperature dependence of the (a) dielectric constant $\varepsilon'$ and (b) dissipation factor $D$ of BINOL$\cdot$2DMSO. The electric field direction and frequency are denoted in the caption.
(c) $T_\mathrm{c1}$ and $T_\mathrm{peak}$ as a function of frequency.
}
\end{figure}

Figure \ref{fig:eps}(a) shows the temperature dependence of the dielectric constant $\varepsilon^{\prime}$ at various frequencies. 
A clear anomaly is observed at $T_\mathrm{c1}=190$ K, indicating a ferroelectric or antiferroelectric phase transition.
Slightly below $T_\mathrm{c1}$, a broad peak in the dissipation factor $D$ is observed at a certain temperature $T_\mathrm{peak}$ [Fig. \ref{fig:eps}(b)] probably due to the domain dynamics \cite{PhysRevB.55.16159}. 
The frequency dependencies of $T_\mathrm{c1}$ and $T_\mathrm{peak}$ with the error bar of FWHM are plotted in Fig.~\ref{fig:eps}(c). 
$T_\mathrm{peak}$ shifts from 190 K to 120~K towards a low-frequency limit, while the transition temperature $T_\mathrm{c1}$ is frequency independent. 
This frequency dependence indicates that the domain motion is frozen at around 120~K. 
Noteworthy, a small kink is observed in the dielectric constant at $T_\mathrm{c2} \sim 120$~K measured at 1~kHz, which indicates the freezing temperature of domains or another phase transition.

\begin{figure}[p]
\centering
\includegraphics[width=7cm]{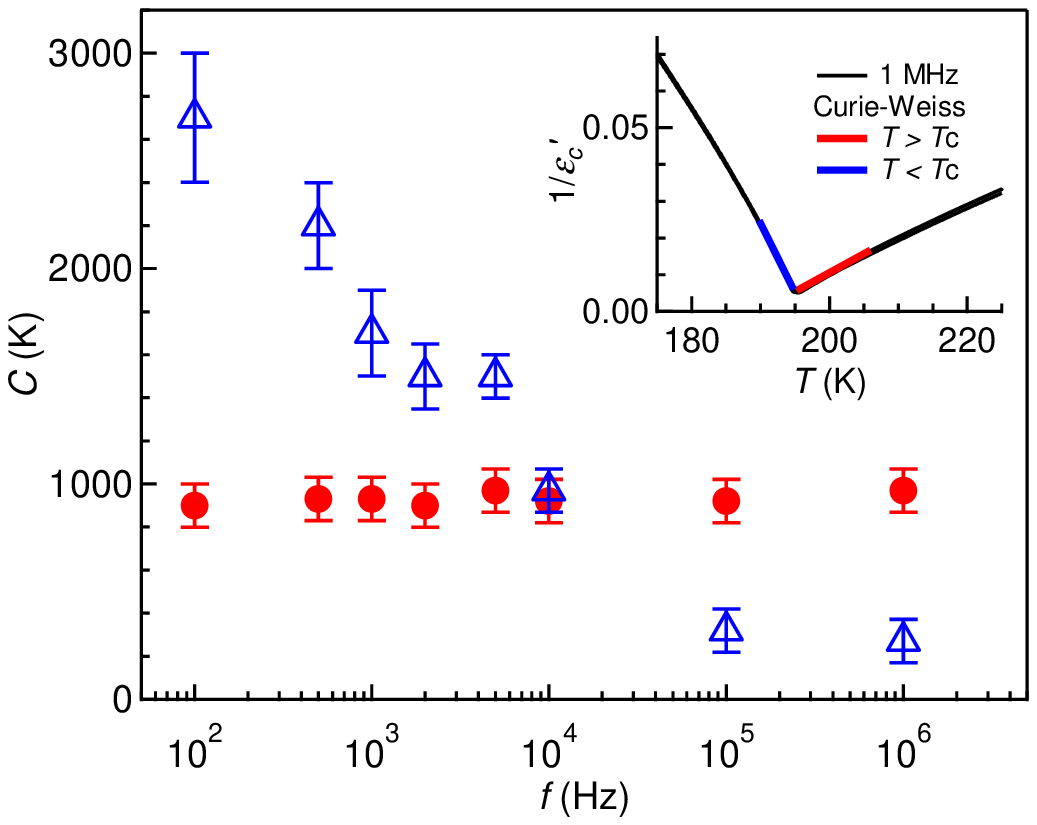}
\caption{\label{fig:cw}
(Color online) Frequency dependence of the Curie constant $C$ for $E||c$. Circles and triangles represent the value above and below $T_\mathrm{c1}$, respectively.
The inset shows the Curie-Weiss fit of $1/\varepsilon'$ for $T>T_\mathrm{c1}$ (red) and for $T<T_\mathrm{c1}$ (blue).
}
\end{figure}

The inset of Fig. \ref{fig:cw} shows the inverse dielectric constant 1/$\varepsilon^{\prime}$ measured at 1~MHz. 
Blue (red) line shows the fitting result below (above) $T_\mathrm{c1}$ based on the Curie-Weiss law 
\begin{equation}
1/{\varepsilon}^{\prime} = |T-T_{\mathrm{c1}}|/C,
\label{eq:CW}
\end{equation}
where $C$ is the Curie constant. 
Both curves below and above $T_\mathrm{c1}$ are well fitted by Eq.~(\ref{eq:CW}).
However, ${\varepsilon}^{\prime}(T)$ below $T_\mathrm{c1}$ deviates from the Curie-Weiss law at frequencies lower than 10~kHz [the hump around 160 K in Fig. \ref{fig:eps}(a)].
The frequency dependence of the Curie constant is plotted in Fig. \ref{fig:cw}.
$C(T>T_\mathrm{c1})$ is almost frequency independent, while $C(T<T_\mathrm{c1})$ decreases at higher frequencies, reflecting the domain-wall dynamics.
At a high-frequency limit, $C(T>T_\mathrm{c1})/C(T<T_\mathrm{c1}) \sim 3.8$, which is close to the Landau-theory value of 2.
The estimated Curie constant around 1000 K is in a typical energy scale for ferroelectrics driven by the order-disorder mechanism, NaNO$_2$\cite{doi:10.1143/JPSJ.16.2207} and tri-glycine sulfate (TGS) \cite{PhysRev.107.1255}.

\begin{figure}[p]
\centering
\includegraphics[width=6.5cm]{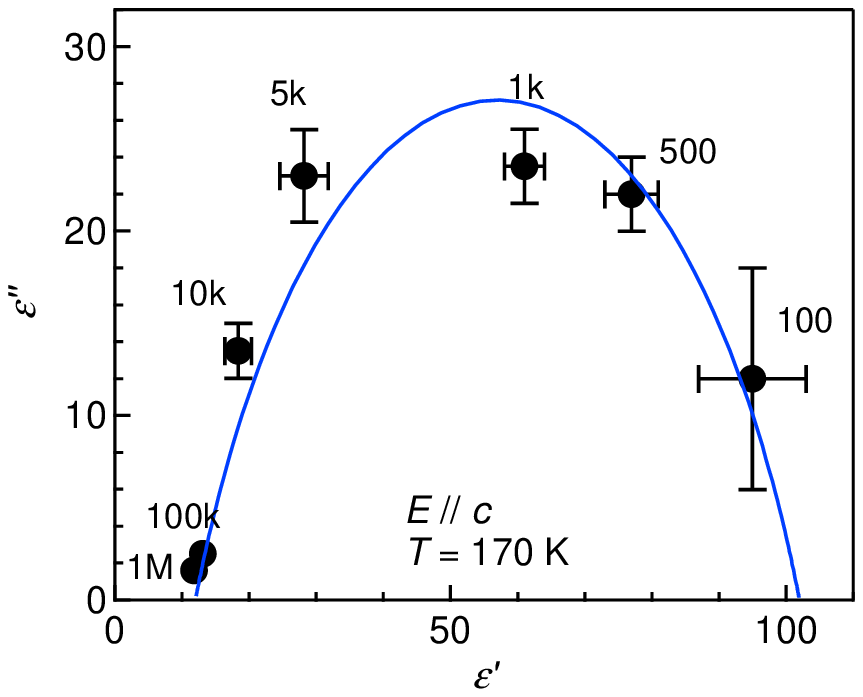}
\caption{\label{fig:cole}
(Color online) Cole-Cole diagram of the complex dielectric constant at 170 K with applying fields along the $c$ axis.
Frequency in the unit of Hz is denoted for each plot.
Cole-Cole law [Eq. (\ref{eq:cole}), with $\alpha=0.3$] is shown by the blue curve for eye guide.
}
\end{figure}

Figure \ref{fig:cole} shows the Cole-Cole diagram of the complex dielectric constant $\varepsilon^\ast = 
\varepsilon^\prime -i\varepsilon^{\prime \prime}$ at 170 K.
The frequency dependence of the real and imaginary parts of the dielectric constant shows a single-dome behavior, indicating only one relaxation mechanism is relevant in this frequency range.
The blue curve shows the Cole-Cole law \cite{doi:10.1143/JPSJ.16.2207}
\begin{equation}
\frac{{\varepsilon}^{\ast}-{\varepsilon}^0}{{\varepsilon}^0-{\varepsilon}^{\infty}}
=\frac{1}{1+(2{\pi}f{\tau}i)^{1-\alpha}},
\label{eq:cole}
\end{equation}
where $\varepsilon^0$ and $\varepsilon^\infty$ are the dielectric constants at low and high frequency limits, and $\tau$ is the relaxation time.
$\alpha$ is a deviation parameter from the Debye model.
With the parameters of $\alpha=0.3$ and $\tau=5$ ms, the experimental plots are reasonably fitted.
The relaxation time exponentially increases towards 120 K as seen from the temperature dependence of $T_\mathrm{peak}$ [Fig.~\ref{fig:eps}(c)].

\begin{figure}[p]
\centering
\includegraphics[width=9cm]{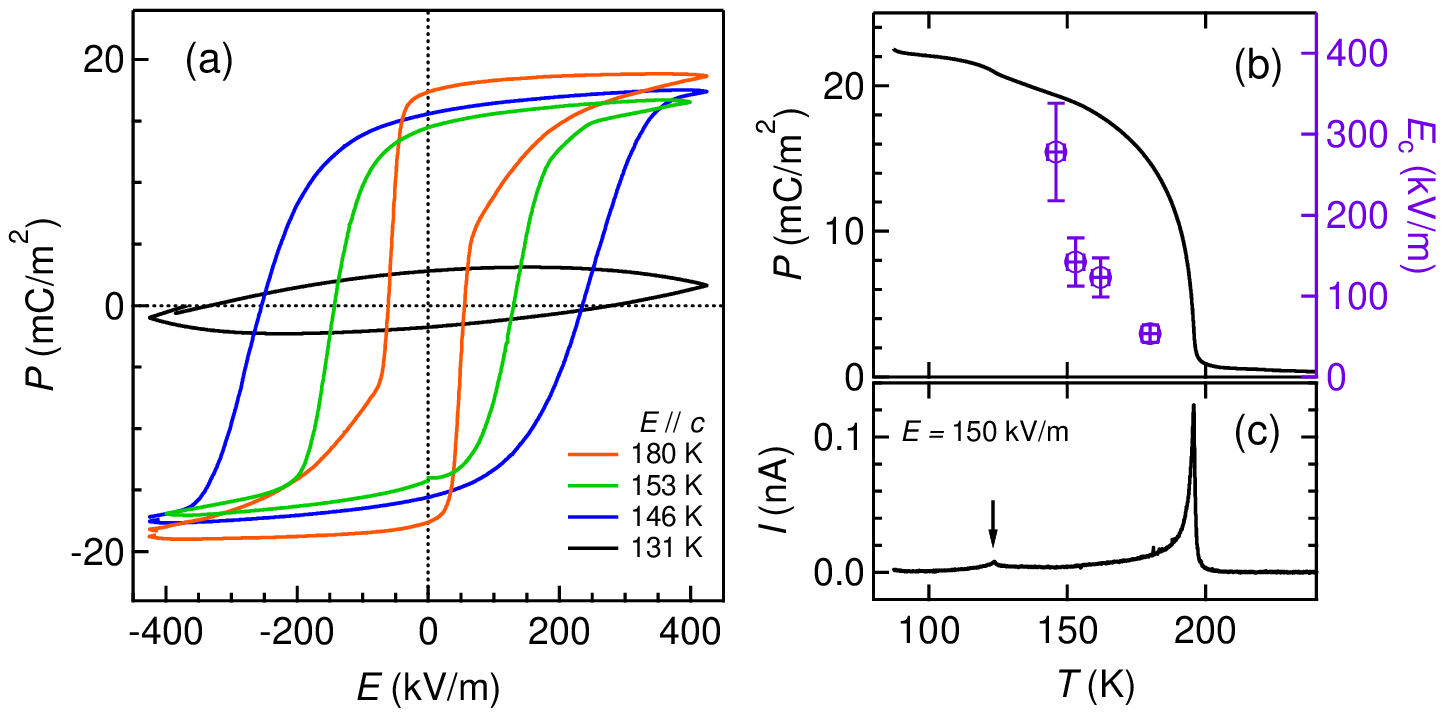}
\caption{\label{fig:pe}
(Color online) (a) $P-E$ hysteresis loops for temperatures below $T_\mathrm{c1}$.
The obtained coercive field $E_\mathrm{c}$ is plotted in Fig.~\ref{fig:pe}(b) right axis.
(b) Polarization $P$ as a function of temperature obtained by integrating the (c) pyrocurrent $I$.
The bias field $E=150$ V/m is along the $c$ axis.
}
\end{figure}

Next, we show the polarization versus electric field ($P-E$) curves measured at the frequency of 5~mHz [Fig. \ref{fig:pe}(a)]. 
Single hysteresis loops are observed, meaning that the phase below $T_\mathrm{c1}$ is ferroelectric, not antiferroelectric. 
The temperature dependence of the coercive field $E_\mathrm{c}$ is plotted in Fig. \ref{fig:pe}(b) right axis.
$E_\mathrm{c}$ nonlinearly increases towards low temperatures and goes over 400~kV/m already at 131~K. 
The uncommon shape of the hysteresis loop for 180 K [orange curve in Fig. \ref{fig:pe}(a)] might be related to the domain-wall motion observed in the dielectric measurements (Fig. \ref{fig:eps}).

Figure \ref{fig:pe}(c) shows the result of pyrocurrent measurement with the temperature sweep rate of $\sim+0.1$ K/s. 
In addition to the clear anomaly at $T_\mathrm{c1}=190$ K, a sharp peak is observed at $T_\mathrm{c2}=125$ K, indicating another phase transition. 
By integrating $I(t)$ and assuming no spontaneous polarization at room temperature, the polarization as a function of temperature is obtained as Fig. \ref{fig:pe}(b). 
The polarization starts to increase at $T_\mathrm{c1}$ and almost saturates below $T_\mathrm{c2}$ at the value of $\sim 20$ mC/m$^2$. 
This polarization value is again comparable to other ferroelectric materials driven by the order-disorder transition of polar molecules \cite{doi:10.1143/JPSJ.16.2207,PhysRev.107.1255}.

\subsection{Specific heat}
\begin{figure}[p]
\centering
\includegraphics[width=7.2cm]{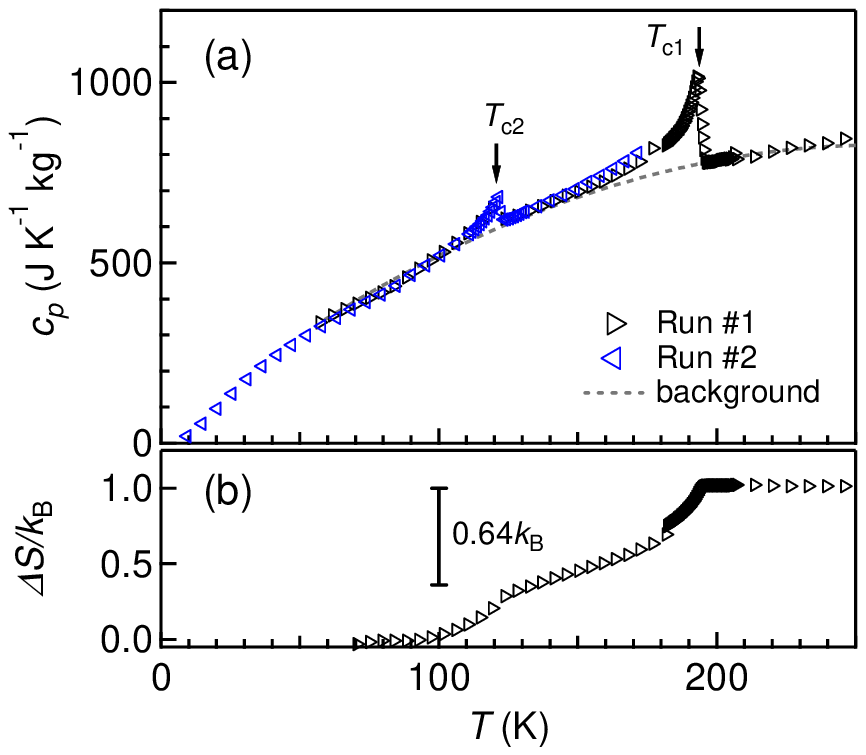}
\caption{\label{fig:cp}
(Color online) (a) Specific heat $c_p$ and (b) entropy change $\Delta S$ per one DMSO as a function of temperature.
The phonon contribution is subtracted as the gray curve in Fig.~\ref{fig:cp}(a).
}
\end{figure}

Figure \ref{fig:cp}(a) shows the specific heat as a function temperature. 
Two sets of results coincide well, including the anomalies at $T_\mathrm{c1}=190$ K and $T_\mathrm{c2}= 125$ K. 
The $\lambda$-shaped anomalies suggest that these transitions are of second-order.
Indeed, all other measurements do not show sizable hysteresis.
The entropy change related to these transitions $\Delta S$ are estimated in Fig.~\ref{fig:cp}(b).
Here, the phonon contribution is subtracted as the gray curve in Fig. \ref{fig:cp}(a), and the entropy is normalized for one DMSO molecule.
The relatively large $\Delta S$ at $T_\mathrm{c1}$ indicates that the ferroelectric transition is driven by the order-disorder mechanism \cite{doi:10.1143/JPSJ.20.2180}.

Here, we discuss the theoretical entropy change for an order-disorder transition of DMSO molecules in this system. 
Previous structural analysis suggests that there are two molecular alignments for each DMSO molecule at room temperature, with the configurational probabilities of major:minor $\sim$ 2/3:1/3 [Fig. \ref{fig:crystal}(b, c)] \cite{doi:10.1021/cg1004648}. 
The minor configuration is preferred for the hydrogen bonding between DMSO and BINOL.
However, in this packing geometry, the DMSO molecule is subjected to larger stress from the BINOL framework, observed as the stretching of the S-C bonding.
By assuming this configurational degree of freedom freezes at $T_\mathrm{c1}$, the expected entropy change per one DMSO is calculated as $\Delta S = (1/3\mathrm{ln}(1/3) + 2/3\mathrm{ln}(2/3))k_\mathrm{B} = 0.64k_\mathrm{B}$.
This value is comparable to the entropy change from 120 K to 190 K, indicating that the fluctuation of the DMSO molecules gradually freezes towards $T_\mathrm{c2}$.
This picture is in line with the temperature dependence of $T_\mathrm{peak}$, which also suggests that the dynamics freezes towards $T_\mathrm{c2}$.

Here, we note on the possible error range of our results.
We notice that the sample becomes opaque and the mass of the sample slightly decreases after measuring the specific heat in a vacuum environment.
This indicates that the guest DMSO molecules can escape from the crystal with the help of vacuum.
We estimate that possible errors of the heat capacity and entropy change are up to $\pm$~5\% and $\pm$~10\%, respectively.
The estimated error does not affect the above discussions.

\subsection{Ultrasound velocity}
\begin{figure}[p]
\centering
\includegraphics[width=7.1cm]{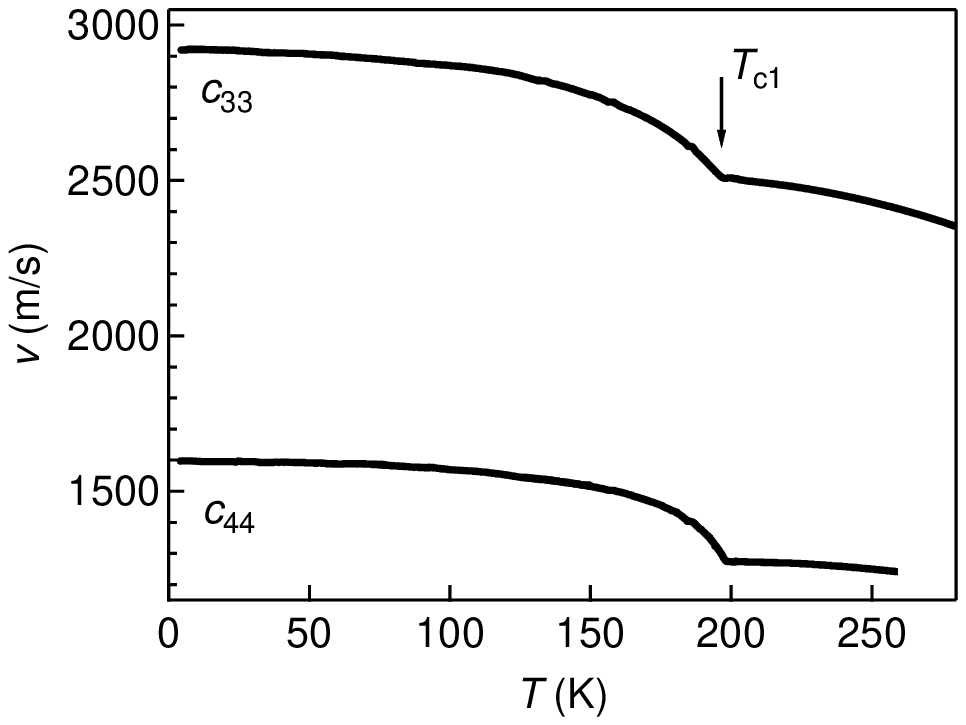}
\caption{\label{fig:sound}
Sound velocity of the $c_{33}$ (38~MHz) and $c_{44}$ (18~MHz) acoustic modes as a function of temperature.
}
\end{figure}

Figure \ref{fig:sound} shows the temperature dependence of the sound velocity of the $c_{33}$ and $c_{44}$ acoustic modes. 
A clear anomaly is observed at $T_\mathrm{c1} = 190$ K, which is consistent with other measurements. 
The phase transition at $T_\mathrm{c2}$ is not detected with these acoustic modes, indicating that the corresponding strains $\epsilon_{zz}$ and $\epsilon_{zx}$ ($\epsilon_{yz}$) are not involved at this transition. 
The slope change at $T_\mathrm{c1}$ is reminiscent of the change of the polarization in Fig. \ref{fig:pe}, indicating that the developing dipole-dipole interaction contributes to the stiffness.
We note that the ultrasound frequency of 10--100~MHz is much higher than the relaxation time estimated by the Cole-Cole plot ($\sim5$~ms), and the effect of domain dynamics is not observed in the sound velocity.

We should note that the echo signal becomes discontinuously worse below 150 K. 
This is probably related to the domain formation below the ferroelectric transition and sound scattering at the domain boundaries. 
The curves in Fig. \ref{fig:sound} show the results obtained by two or three runs of measurements, which are well reproduced.
The effect of the domain boundaries is observed in the optical spectroscopy as well.

\subsection{Optical spectroscopy}
\begin{figure}[p]
\centering
\includegraphics[width=8.6cm]{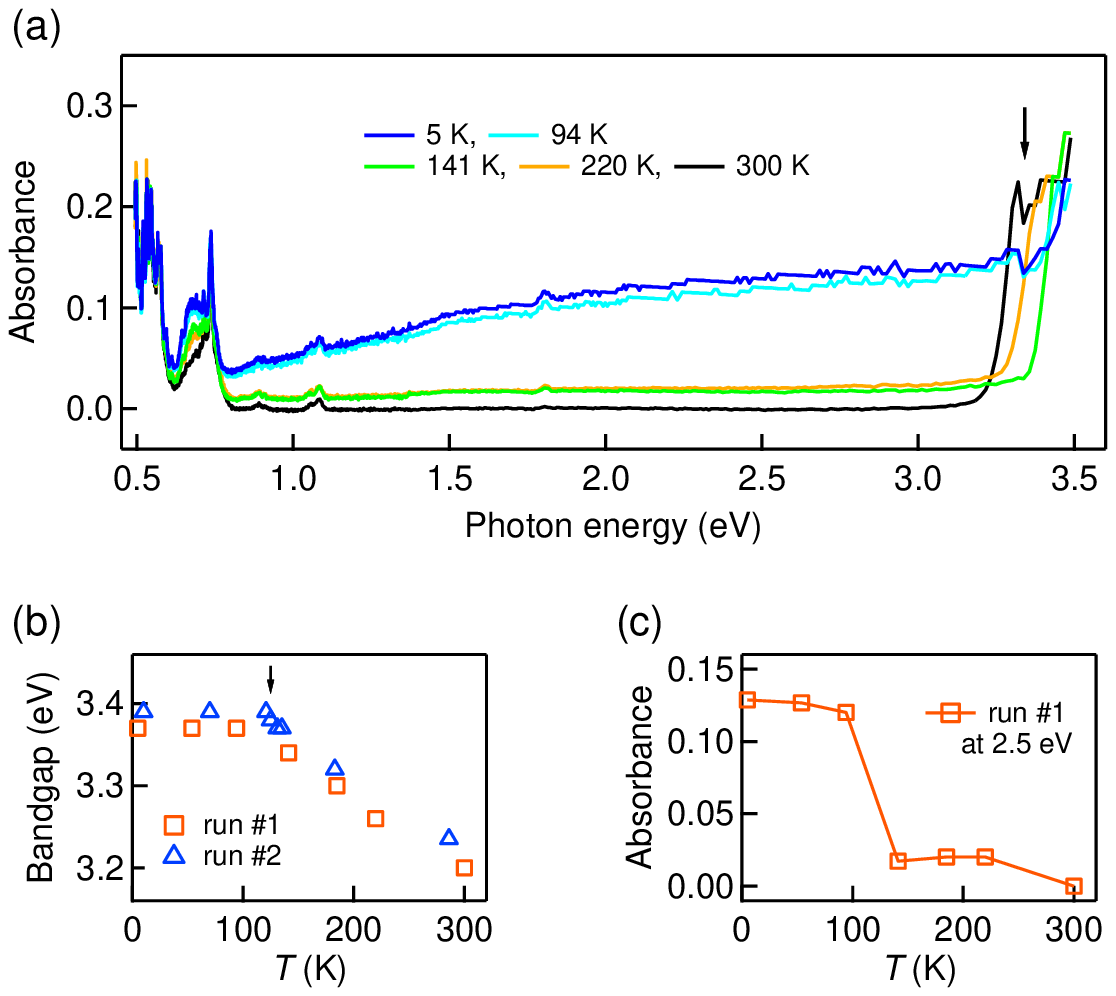}
\caption{\label{fig:abs}
(Color online) (a) Absorbance spectra at various temperatures. The arrow indicates the band edge.
(b) Temperature dependence of the bandgap. The arrow indicates the change of slope at $T_\mathrm{c2}$.
(c) Temperature dependence of the absorbance at 2.5 eV.
}
\end{figure}

Figure \ref{fig:abs}(a) shows the absorption spectra of the measurement run \#1 (2.6 mm thickness) at temperatures from 300 K to 5 K. 
The absorbance at 300 K in the visible range is assumed to be zero \cite{https://doi.org/10.1002/cphc.201800950} and the background due to light scattering is subtracted. 
The absorption peaks below 1 eV come from the intramolecular vibrational modes and their higher harmonics \cite{molecules21111541,Wallace2015}, which are almost temperature independent. 
The absorption edge at 3.3 eV corresponds to the bandgap, consisting mainly of the $\pi-\pi^*$ transitions of BINOL \cite{doi:10.1021/cg1004648}.
The temperature dependence of the bandgap is plotted in Fig.~\ref{fig:abs}(b). 
The bandgap increases towards lower temperatures and saturates at $T_\mathrm{c2}$.
This probably reflects the change of the packing structure which affects the intermolecular hybridization \cite{PhysRevB.101.165102,singleton}.

The absorbance change at 2.5 eV, where no optical absorption exists, is plotted in Fig.~\ref{fig:abs}(c).
The drastic change occurs at around 120~K, where the absorbance increases in all energy ranges. 
By eye observations, we find that the crystal becomes opaque and the light scattering increases at this temperature. 
The absorbance tends to be larger in the higher photon energy (shorter wavelength), indicating Rayleigh scattering \cite{doi:10.1080/00150193.2011.577673}. 
Based on the other measurements above, we speculate that this light scattering is due to crystallographic domains, probably related to the structural transition from tetragonal to lower-symmetry phase (orthorhombic or monoclinic) at $T_\mathrm{c2}$.
The mismatch of the refractive index at the domain boundaries causes strong light scattering \cite{PhysRevLett.112.247201}.
In contrast, the light scattering does not change at $T_\mathrm{c1}$.
This indicates that the ferroelectric transition at $T_\mathrm{c1}$ does not form domain boundaries by lowering the crystallographic symmetry from tetragonal.

\subsection{Powder XRD}
We have performed the low-temperature powder XRD measurements at 220, 150, 100, and 50 K.
The full diffraction patterns are presented in Supplemental Material (SM) \cite{suppl}.
The diffraction pattern at 220 K can be reasonably fitted by the reported crystal structure at room temperature \cite{doi:10.1021/cg1004648}.
Subsequent symmetry lowering is observed in the low-temperature results.

\begin{figure}[p]
\centering
\includegraphics[width=8.6cm]{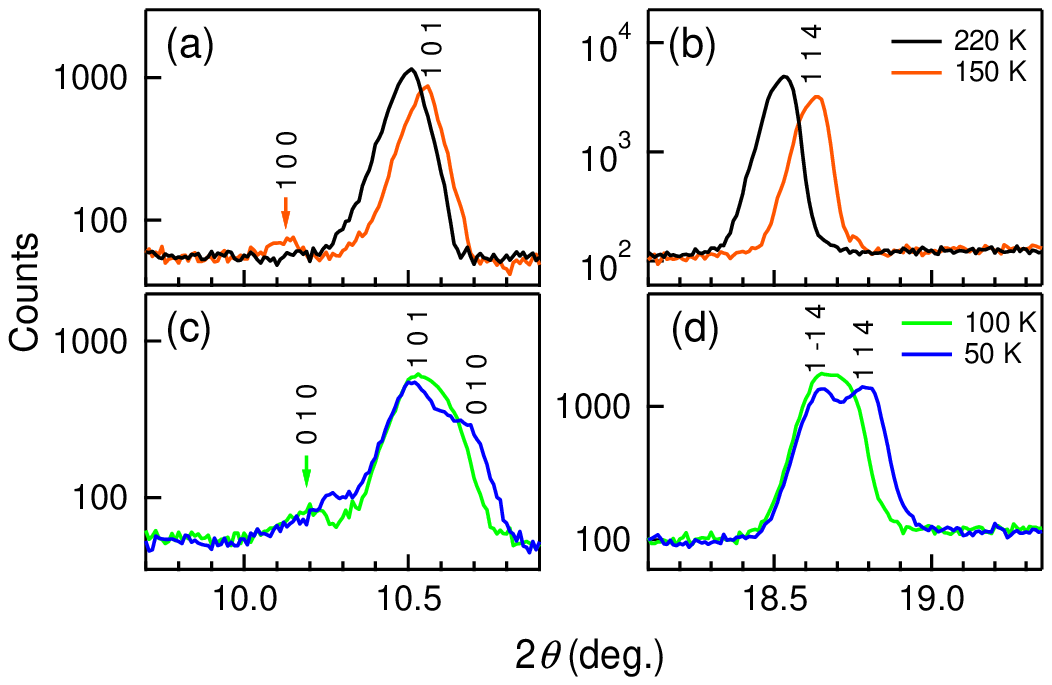}
\caption{\label{fig:101}
(Color online) Enlarged powder XRD patterns of the (a, b) tetragonal and (c, d) monoclinic phases. For full XRD patterns, see Supplemental Material \cite{suppl}.
}
\end{figure}

\begin{figure}[p]
\centering
\includegraphics[width=8cm]{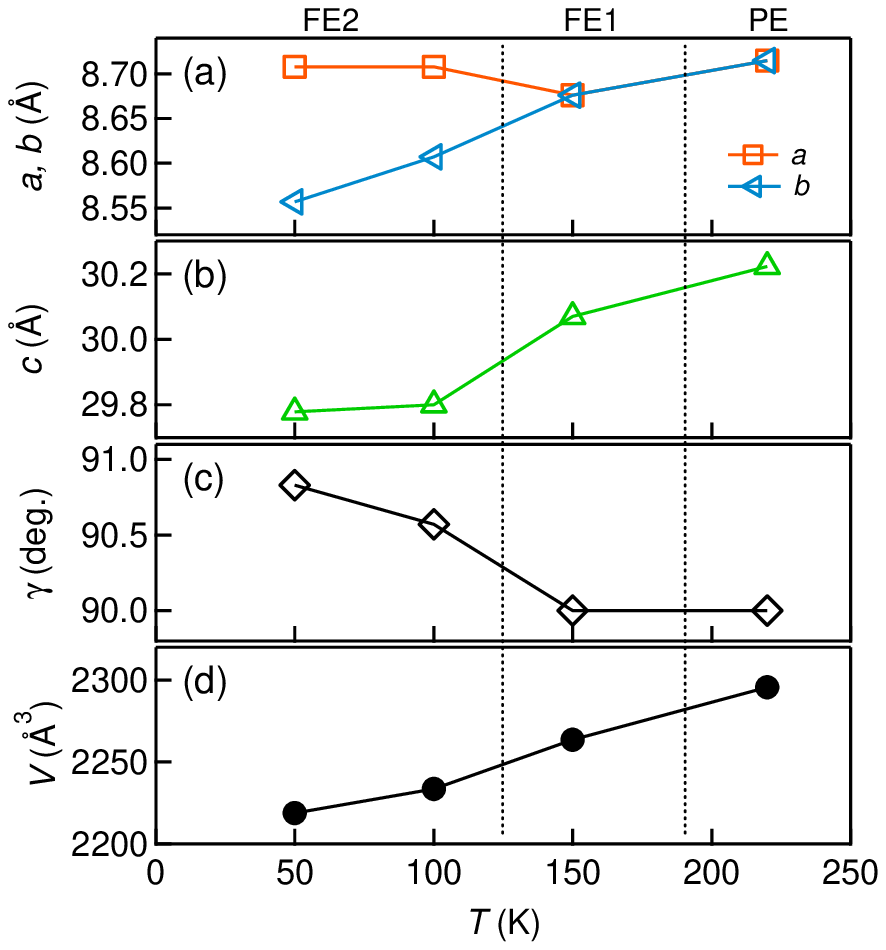}
\caption{\label{fig:lattice}
(Color online) Temperature dependence of the unit cell parameters. (a, b) Lattice constants, (c) monoclinic angle, and (d) unit cell volume.
}
\end{figure}

Figures \ref{fig:101}(a-d) show the diffraction patterns around the $101$ and $114$ peaks.
The $101$ and $114$ peaks do not split at 150 K [Figs. \ref{fig:101}(a, b)], however, an additional peak appears at 10.2 deg. at 150~K.
This indicates that the transition at $T_\mathrm{c1}$ changes the structure from higher-symmetry tetragonal to lower-symmetry tetragonal ones.
In the low temperature phase below $T_\mathrm{c2}$ [Figs. \ref{fig:101}(c, d)], clear splittings of $101$ and $114$ peaks are observed, indicating that the symmetry changes from tetragonal to monoclinic (or triclinic).
Indeed, the overall diffraction patterns of the low-$T$ phase can be fitted by a monoclinic structure (unique axis $c$).
Figures~\ref{fig:lattice}(a-d) summarize the lattice constants, monoclinic angle, and unit-cell volume.
For convenience, we call the tetragonal paraelectric phase as PE, the tetragonal ferroelectric phase as FE1 ($T_\mathrm{c2}<T<T_\mathrm{c1}$), and the monoclinic ferroelectric phase as FE2 ($T<T_\mathrm{c2}$).

Next, we discuss the space groups of FE1 and FE2.
The candidate space groups can be significantly reduced to the product set of the chiral and polar space groups, lacking the mirror and inversion symmetries, respectively.
$P4$, $P4_1$, $P_2$, $P4_3$, $I4$, and $I4_1$ are the candidate space groups for the ferroelectric tetragonal structure (FE1), while $P112$, $P112_1$, and $A112$ are the candidates for the ferroelectric monoclinic structure (FE2), satisfying these conditions.
It is notable that the orthogonal structures do not satisfy both chiral and polar conditions simultaneously.
From these candidates, we narrow down the space group based on the extinction rule.

Figure \ref{fig:101}(a) indicates that FE1 shows a $100$ peak which is canceled in PE.
The appearance of this index narrows the candidates down to $P4$, $P4_1$, $P_2$, $P4_3$.
Among the $00l$ peaks, only the $004$ and $008$ peaks are observed for PE1, indicating $4_1$ ($4_3$) screw axis along the $c$ axis (see SM \cite{suppl}).
Thus, we conclude that the space group of FE1 is $P4_1$ ($P4_3$).
These are the maximal subgroups from $P4_12_12$ ($P4_32_12$) of PE.
The active irreducible representation (IR) describing the crystal-symmetry breaking is $A_2$.
Figure \ref{fig:101}(c) indicates that PE2 shows the $010$ peak, but not the $100$ peak expected at 10.06 deg.
Among the candidate space groups, only $P112_1$ satisfies this condition.
Thus, we conclude the space group of FE2 as $P112_1$.
Again, $P112_1$ is the maximal subgroup from $P4_1$ ($P4_3$) with the active IR of $B$.
The group-subgroup relationship is consistent with the observed second-order transitions at $T_\mathrm{c1}$ and $T_\mathrm{c2}$.

\section{Discussion}
Here, we discuss the consistency of the experimental results.
The ultrasound velocity clearly detects the transition at $T_\mathrm{c1}$, while does no the one at $T_\mathrm{c2}$.
This reflects the soft mode at $T_\mathrm{c2}$.
From FE1 to FE2, $\epsilon_{xy}$ and $\epsilon_\mathrm{T}=(\epsilon_{xx}-\epsilon_{yy})/2$ belonging to the IR $B$ are the active strains.
The expected acoustic soft modes are the $c_{66}$ and $c_\mathrm{T}=(c_{11}-c_{12})/2$.
In this study, we have only measured the $c_{33}$ and $c_{44}$ modes because of the crystal habit.
Therefore, the insensitivity at $T_\mathrm{c2}$ is consistent with the symmetry arguments.

Next, we discuss the results of the optical spectroscopy.
The drastic increase of light scattering is observed at $T_\mathrm{c2}$, while it does not change at $T_\mathrm{c1}$.
The difference indicates that the crystallographic domain is the origin of the light scattering.
At $T_\mathrm{c2}$, the unit cell deforms from tetragonal to monoclinic, leading to two domain structures with the shear strains.
Such domain boundaries cause the refractive-index mismatch and lead to the opaque appearance of the crystal.
A similar effect, the decrease of the acoustic signal, is observed in the ultrasound measurement as well.
On the other hand, the transition at $T_\mathrm{c1}$ does not change the tetragonal unit cell, although the ferroelectric domain structures are formed as indicated by the hysteresis loop [Fig.~\ref{fig:pe}(a)].
Such ferroelectric domains would also be observed by using a polarizing microscope.

Here, we comment on the order-disorder transition of DMSO molecules at $T_\mathrm{c1}$.
With the symmetry lowering from $P4_12_12$ to $P4_1$, $2_1$ and $2$ axes in the $a$-$b$ plane are lost.
Because of this symmetry lowering, the nearest neighbor DMSO sites [Fig. \ref{fig:crystal}(b, c)] become nonequivalent.
Most probably, at $T_\mathrm{c1}$, one DMSO site settles in the major configuration and another in the minor one.
Detailed studies on the crystal structures at low temperatures are left for future works.

Last, we comment on the possible formation of electric Bloch skyrmions in this compound.
The point group changes from $422$ to $4$ at $T_\mathrm{c1}$ satisfies the necessary condition to realize electric Bloch skyrmions in a bulk system \cite{PhysRevB.102.024110}.
The strange hysteresis curve in the $P-E$ curve [Fig.~\ref{fig:pe}(a), 180 K] and the broad peak of $D$ [Fig.~\ref{fig:eps}] might be related to this kind of ferroelectric domains.

\section{Conclusion}
We have characterized the physical properties of a chiral organic compound BINOL$\cdot$2DMSO.
We have found two second-order phase transitions at $T_\mathrm{c1}=190$ K and $T_\mathrm{c2}=125$ K, where the space group changes as $P4_12_12$ ($P4_32_12$) $\rightarrow$ $P4_1$ ($P4_3$) $\rightarrow$ $P112_1$.
The transition at $T_\mathrm{c1}$ is characterized by the spontaneous polarization driven by the order-disorder transition of the DMSO molecules.
At $T_\mathrm{c2}$, the crystal symmetry is further lowered by the monoclinic distortion.
Because of the domain boundaries, the transmission of light and acoustic waves is strongly disturbed in the monoclinic phase.

In this paper, we report a new member of the chiral organic ferroelectrics.
Although the first ferroelectricity had been reported on a chiral system (Rochelle salt), chiral ferroelectrics in purely organic systems are still seldom \cite{Horiuchi2008,li2022highest,Li5878,acsnano7b07090}.
Besides, most of the reported chiral organic ferroelectrics show hysteretic behavior in temperature dependence.
BINOL$\cdot$2DMSO, which shows a second-order ferroelectric transition, could be an attractive platform to deepen the understanding of organic ferroelectrics.

\begin{acknowledgment}
We thank M. Akatsu for sharing the ultrasound measurement software. 
This work was partly supported by JSPS KAKENHI, Grant-in-Aid for Scientific Research (No. 20K14403).

\end{acknowledgment}

\bibliographystyle{jpsj}
\bibliography{b1}

\end{document}